\documentclass[preprint]{vgtc}

\preprinttext{}

\graphicspath{{figures/}{pictures/}{images/}{./}}

\usepackage{times}

\usepackage{booktabs}
\usepackage{multirow}
\usepackage{array}
\usepackage{enumitem}
\usepackage{amsmath, amssymb}
\usepackage[table]{xcolor}
\usepackage{balance}
\usepackage{mathptmx}
\usepackage{xcolor}
\usepackage{xspace}
\usepackage{cite}
\usepackage{hyperref}
\usepackage{cleveref}

\newcommand{\rev}[1]{\textcolor{black}{#1}}

\newcommand{\osf}{\href{https://osf.io/23r67}{OSF (osf.io/23r67)}}

\definecolor{scopeElement}{HTML}{993C1D}
\definecolor{scopeBundle}{HTML}{0F6E56}
\definecolor{scopeGlobal}{HTML}{185FA5}
\definecolor{scopeMulti}{HTML}{854F0B}

\newcommand{\sElement}{\textcolor{scopeElement}{Element}\xspace}
\newcommand{\sBundle}{\textcolor{scopeBundle}{Bundle}\xspace}
\newcommand{\sGlobal}{\textcolor{scopeGlobal}{Global}\xspace}
\newcommand{\sMulti}{\textcolor{scopeMulti}{Multi-view}\xspace}

\title{A Task Taxonomy for Edge and Trail Bundling}

\author{Markus Wallinger\thanks{e-mail: markus.wallinger@tum.de}\\ %
        \scriptsize TU Munich %
\and Stephen  Kobourov\thanks{e-mail: stephen.kobourov@tum.de}\\ %
     \scriptsize TU Munich}

\teaser{
  \centering
  \includegraphics[width=0.95\linewidth]{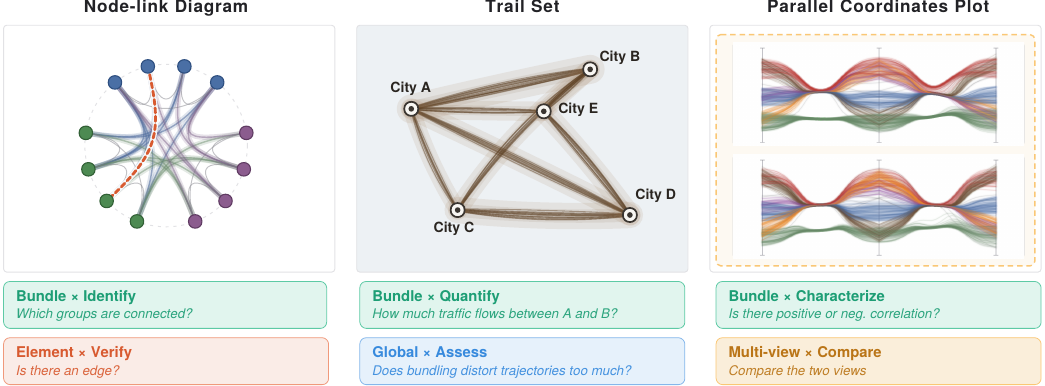}
  \caption{Edge bundling illustrated on three representation types, each supporting different tasks from our Scope~$\times$~Action taxonomy.
  (a)~A node-link diagram: bundling reveals which groups of modules are strongly connected (\sBundle~$\times$~Identify) but makes it harder to verify whether two specific modules share a direct dependency (\sElement~$\times$~Verify).
  (b)~A geographic trail set: bundling supports estimating traffic volume between cities (\sBundle~$\times$~Quantify) but raises the question of whether routes have been shifted away from their true paths (\sGlobal~$\times$~Assess).
  (c)~A bundled parallel coordinate plot: bundling makes it possible to see whether variables increase or decrease together (\sBundle~$\times$~Characterize) and to compare patterns across two datasets (\sMulti~$\times$~Compare).
  Different colors in the figure correspond to the four different scope levels: \sElement, \sBundle, \sGlobal, and \sMulti.}
  \label{fig:teaser}
}

\abstract{
Edge bundling reduces visual clutter by aggregating similar edges, yet practitioners lack a structured vocabulary for reasoning about the tasks that bundled visualizations support.
Such a vocabulary is needed both to evaluate the general utility of bundling and to compare different bundling approaches.
We address this gap by assembling a corpus of 102 papers, 49 of which contain explicit bundling tasks, spanning node-link diagrams, geographic trail sets, and parallel coordinate plots.
From this corpus, we derive a task taxonomy organized as a matrix of \emph{scope} (Element, Bundle, Global, Multi-view) crossed with \emph{action} (Verify, Identify, Characterize, Quantify, Compare, Assess), instantiated across the three representation types.
We show that bundling simultaneously \emph{enables} tasks (bundle-level and global reasoning) and \emph{disables} others (element-level precision), a duality not captured by existing task frameworks.
Our coded corpus and taxonomy are released as supplemental material on \osf.
}

\keywords{Task taxonomy, edge bundling, trail bundling, parallel coordinates, network visualization.}

\begin{document}

\firstsection{Introduction}

\maketitle

Edge and trail bundling~\cite{Holten06,HurterET12,ZwanCT16,LhuillierHT17} spatially aggregates similar edges and paths in a drawing, trading visual clutter for overdraw and abstraction.
Bundling techniques apply to node-link diagrams, geographic trail sets, and parallel coordinate plots~(PCPs) alike.
Yet, \rev{their evaluation has centered largely on algorithmic quality metrics and visual aesthetics~\cite{LhuillierHT17,Wallinger2026Metrics,Saga16} rather than on a systematic understanding of the analytic tasks bundling is meant to support, with perceptual task-based studies remaining rare~\cite{WallingerARPA25}}.

The proposed task taxonomy for bundling serves three main purposes.
First, it provides a basis for \emph{evaluating} whether bundling is useful in a given analytical setting. Second, it makes it possible to compare different bundling methods. %
Third, it  helps \emph{connect} existing bundling quality metrics (e.g., ink reduction, distortion measures, ambiguity) to the tasks they are meant to serve, clarifying which metrics are relevant for which analytical goals.

A 2017 edge bundling survey by Lhuillier~\cite{LhuillierHT17} includes a brief discussion of tasks, mapping bundling capabilities to the general task frameworks of Lee et al.~\cite{LeePPFH06} and Brehmer and Munzner~\cite{BrehmerM13}. The treatment is compact and largely conceptual, and the survey itself concedes that no accepted metrics exist to substantiate claims of task support. We identify three specific shortcomings.

First, existing task discussions fail to account for the fact that bundling introduces a new class of perceptual objects, the \emph{bundles}, which support tasks absent from more general taxonomies.
Identifying the major connection groups in a network, comparing the relative thickness of two bundles, or identifying which edges belong to a bundle make sense only after bundling has been applied.
No prior framework provides vocabulary for these bundle-level tasks.

Second, task semantics shift substantially across representation types.
Tracing in a node-link diagram (following an edge between named endpoints), in a trail set (following a spatial trajectory), and in a PCP (reading a multivariate profile) are fundamentally different operations despite sharing an abstract verb.

Third, prior discussions overstate what bundling supports.
The trade-off between precision and aggregation means that some tasks are \emph{enabled} by bundling (bundle-level and global reasoning) while others are \emph{disabled} by it (individual element tracing and attribute reading).
Our corpus analysis confirms that tasks such as characterizing or comparing individual elements are missing from the bundling literature -- not because they are meaningless, but because bundling makes them impractical.
At the same time, tasks that bundling clearly enables, such as global faithfulness assessment, were absent from prior task discussions.

We address these gaps through a systematic coding study.
We assemble a corpus of 102 bundling papers spanning node-link diagrams, trail sets, and PCPs. 
Unlike prior work, which maps bundling \emph{onto} existing generic frameworks, we derive a task vocabulary \emph{from} bundling practice: from how tasks are actually stated, evaluated, and implied across the literature.
From this corpus, we derive a two-dimensional taxonomy of \emph{scope}~$\times$~\emph{action} (verify, identify, characterize, quantify, compare, assess), instantiated across the three representation types; see \cref{fig:teaser} for examples.

\section{Background}
\label{sec:background}

\paragraph{Bundling across Representation Types.}

Bundling operates on three data types with different semantics.

A \textbf{node-link diagram} maps an abstract graph $G{=}(V,E)$ to the plane with a layout algorithm determining node positions and edge routing.
Bundling modifies edge routing, but preserves node positions and the underlying relational data~\cite{lhuiller_2017,epb}.

A \textbf{trail set} is a collection of spatially embedded trajectories~\cite{HurterTC09} (flights, eye tracking, vehicle paths).
Here, the geometry \emph{is} the data: deforming a trail through bundling distorts real spatial information, making faithfulness assessment qualitatively different from the graph case.
However, bundling does not alter the underlying origin-destination data or the recorded positions; it changes only the visual routing between them.

A \textbf{parallel coordinate plot}~(PCP) represents multivariate records as polylines crossing parallel axes.
Bundling aggregates records with similar value profiles~\cite{HeinrichW15,McDonnellM08,PalmasW16}.
The ``edges'' here are data records, not relationships.
As in the trail case, bundling modifies the visual routing of polylines but does not alter the underlying data values at the axes; the distortion occurs in the inter-axis space.

In all three cases, bundling changes \emph{how} elements are drawn without changing \emph{what} they represent.
The severity of the resulting visual distortion, however, differs across representation types.
We use the term \emph{element} to refer to the atomic unit in each context: an edge in a graph, a trail in a trajectory set, or a polyline in a PCP.

\paragraph{Task Frameworks.}

Our taxonomy builds on Lee et al.'s graph task taxonomy~\cite{LeePPFH06}, Brehmer \& Munzner's multi-level typology~\cite{BrehmerM13}, the clutter-reduction taxonomy of Ellis \& Dix~\cite{EllisD07}, and group-level graph visualization taxonomies~\cite{SaketSKB14, VehlowBW17}, which introduced tasks for reasoning about \emph{groups} of nodes and edges.
Bundle-level reasoning is closer to this group-level analysis than to the node- and edge-level focus of Lee et al.

The Scope--Action--Target~(SAT) framework of Oddo et al.~\cite{Oddo2026} provides the closest structural template, organizing tasks into triplets of scope (the scale of data considered), action (a verb defining the goal), and target (the data entity of interest).
Our framework adopts SAT's scope and action dimensions but differs in the role of the target.
In the context of unlabelled graphs, the target dimension addresses interesting graph-theoretic structures.
In bundling, the target is simpler: at the element scope, it is always an individual edge, trail, or polyline; at the bundle scope, it is the perceptual aggregate; at the global scope, it is the entire visualization.
Different bundling techniques (directional, density-based, hierarchical) do not constitute distinct targets, since humans reason about bundles as aggregate visual objects regardless of how they were produced.
The variety of what we can  \emph{do} with bundles is captured by the action dimension.
We therefore organize our taxonomy as a \textbf{Scope~$\times$~Action} matrix, instantiated by representation type.

\paragraph{Bundling Methods and Quality Metrics.}

Representative algorithmic families include hierarchical edge bundling~\cite{Holten06}, force-directed bundling~\cite{HoltenW09}, image-based methods~\cite{HurterET12,ZwanCT16,LhuillierHT17}, geometry-based approaches~\cite{CuiZQWL08,lambert_2010}, confluent drawings~\cite{BachRHMD17}, and Edge-Path bundling~\cite{epb,spanepb,ArchambaultLNPT24,WallingerPTALN25}.
For a comprehensive algorithmic treatment, we refer to~\cite{LhuillierHT17}.

Several quality metrics~\cite{Wallinger2026Metrics} have been proposed, including ink reduction~\cite{EllisD06}, edge displacement~\cite{Saga16,epb}, and ambiguity~\cite{epb}.
However, these metrics are typically discussed in isolation from underlying tasks, with few exceptions~\cite{WallingerARPA25}.
Our taxonomy provides a principled way to connect them. For example, ambiguity metrics benefit element- or bundle-level Identify and Verify tasks, and ink reduction metrics serve bundle-level and global-level Identify tasks.

\section{Methodology}
\label{sec:method}

We assembled a corpus of 102 papers by collecting papers cited in a prior bundling survey~\cite{LhuillierHT17}, supplemented by a forward-citation search on bundling techniques.
We included papers that either (a)~propose or evaluate a bundling method, (b)~use bundling in an application, or (c)~conduct user studies involving bundled visualizations.
While this approach does not guarantee exhaustive coverage, it captures the core bundling literature and a broad cross-section of application papers spanning all three representation types.

\paragraph{Coding procedure.}
\rev{Two authors developed the coding scheme on a random subset of 30 papers using open coding, identifying every passage where a paper states, evaluates, or implies a task, and agreeing on a reference set of codes for that subset.}
For each task mentioned, we recorded: the scope level, the action type, whether the mention was \emph{explicit} (stated as a task, evaluated in a study, or listed as a design requirement) or \emph{implicit} (inferred from system design, usage scenario, or figure captions), the representation type, and the paper's original terminology.
After the initial round, we organize tasks along two dimensions:
\begin{itemize}[leftmargin=*, nosep]
  \item \textbf{Scope:} \sElement, \sBundle, \sGlobal, or \sMulti (the perceptual granularity at which attention operates).
  \item \textbf{Action:} verify, identify, characterize, quantify, compare, or assess (the analytical goal).
\end{itemize}
\rev{Because coding proceeded by consensus on this shared reference set rather than by independent double-coding, we do not report a formal inter-rater reliability coefficient; the agreed codes served as the reference standard, and a human coder examined every retained mention to confirm that the raw text and its evidence supported the assigned tag.}

\paragraph{Bundling attribution filter.}
Many papers use bundling as one component within a larger visualization system (e.g., NodeTrix~\cite{HenryFM07} combines adjacency matrices with bundled edges).
We applied a bundling attribution filter during coding: a task was attributed to bundling only if bundled edges or bundling-created aggregates were the visual component that supported or hindered the task.
Tasks performed on non-bundled components (e.g., reading an adjacency matrix cell) were excluded.
This ensures that our corpus analysis reflects what people can do specifically \emph{with bundles}, not what they do with visualization systems that happen to include bundling.

\paragraph{LLM-assisted screening.}
Following recent work showing that LLMs can support qualitative coding when given a clear codebook~\cite{TaiBXS24}, we used an LLM (Claude Sonnet~4.6) as a first-pass screener.
Each paper was submitted with our coding scheme, synonym lists mapping literature phrasings to our canonical labels, and the bundling attribution filter with worked examples; the model returned candidate task mentions with proposed classifications and attribution evidence.
\rev{Then, a human coder reviewed every annotation, correcting approximately 15\% of the LLM's codes. These were mostly over-attributions from related- or future-work passages, duplicates, and scope confusions between \sGlobal and \sMulti (e.g., a cross-view comparison miscoded as \sGlobal~$\times$~Compare), rather than genuine misclassifications. The 15\% reflects the first-pass screening filter, not residual error in the human-verified corpus.}
The prompt is available in the supplemental material.

\section{Task Taxonomy}
\label{sec:taxonomy}

\begin{figure}[t]
    \centering
    \includegraphics[width=0.80\linewidth]{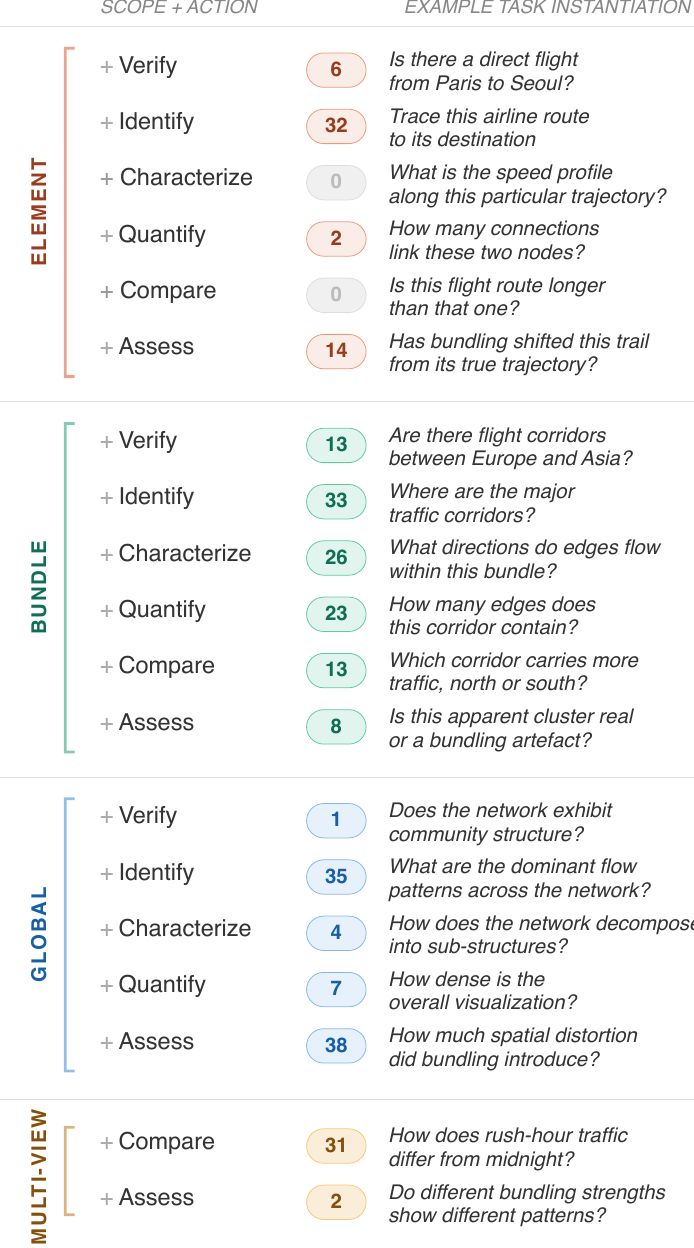}
    \caption{The Scope~$\times$~Action taxonomy. Each row pairs an action type with an example task instantiation. Colored ovals indicate the number of unique papers that mention each combination. Gray ovals mark tasks that bundling hinders: coherent analytical questions that received zero mentions because bundling visually merges the individual elements needed to perform them. Structurally empty cells (e.g., \sMulti~$\times$~Verify) are omitted entirely. The full coded corpus is available in the supplemental material (Table 1).}
    \label{fig:scope_action}
\end{figure}

Our taxonomy is a matrix of four \emph{scope levels} crossed with six \emph{action types}; see \cref{fig:scope_action}.
Each combination represents an abstract task and the scope level determines the target.
Concrete instantiations vary by representation type.
Not every cell is populated: some combinations are \emph{structurally empty}, meaning no meaningful task exists (e.g., \sMulti~$\times$~Verify collapses into within-view verification), while others are \emph{hindered by bundling}, meaning the task is coherent but bundling makes it difficult by merging the elements needed to perform it (e.g., \sElement~$\times$~Characterize).

\subsection{Scope: Four Levels of Attention}
\label{sec:scope}

Bundling produces a hierarchy of perceptual objects that determines where human attention is directed.

At the \textbf{element level}, the focus is on single connections, trails, or records.
Because bundling visually merges elements, tasks at this level often
require mental disentanglement or interaction support such as
lensing~\cite{WongCG03} or selective unbundling~\cite{HurterET11}.
Two element-level tasks, Characterize and Compare, are hindered by bundling: characterizing the profile of one edge or comparing two individual edges requires isolating them from the bundle first, effectively undoing what bundling provides.
Verify and Quantify are also difficult here, as both require disentangling overlapping paths.

At the \textbf{bundle level}, one engages with the perceptual aggregates that bundling creates.
Bundles are emergent visual constructs not present in the raw data as they arise from the algorithmic transformation.
Tasks at this level include identifying the major connection groups or comparing bundle thickness.
These are related to the group-level tasks discussed in \cref{sec:background} but are specific to the bundling context, where the groups are defined by spatial proximity rather than by data attributes.

At the \textbf{global level}, one reasons about the entire visualization to perceive macro-patterns (communities, flows, coverage, anomalies) that would be obscured by clutter in an unbundled view.
\sGlobal~$\times$~Compare is structurally empty: there is only one global view per visualization, so any comparison requires a second view, which falls under the multi-view scope.

At the \textbf{multi-view level}, we compare two or more bundled views showing different datasets, time steps, parameter settings, or spatial extents.
This scope is inherently one of comparison.
One might argue that ``finding the same cluster in both views'' is an Identify task rather than a Compare task, but locating a structure in view~B that was first seen in view~A is operationally a comparison: we need to match across views, which requires assessing similarity.
Verify, Identify, Characterize, and Quantify, when applied across views, all reduce to either a within-view operation at a lower scope or a comparison of what is seen in each view.
We therefore populate only Compare and Assess at this level.
\sMulti~$\times$~Assess captures the specific case where a user evaluates whether the choice of bundling parameters or representation type affects which patterns are visible.

\subsection{Action: Six Question Types}
\label{sec:action}

The action dimension captures the analytical goals. 
We identify six actions, each defined by a distinctive verb.

\begin{description}[leftmargin=0pt, labelindent=0pt, itemsep=2pt, topsep=2pt, parsep=0pt]
  \item[Verify (\emph{Does X exist?})]
  Check for the presence of a specific element, bundle, or structural pattern.
  At the element level, Verify often depends on Identify: confirming that an edge exists typically requires tracing it first.
  This dependency helps explain why Verify appears less frequently than Identify in the corpus.

  \item[Identify (\emph{What or where is X?})]
  Locate a target or read a basic property. This subsumes tracing (following an element) and attribute lookup (reading a single value such as weight).

  \item[Characterize (\emph{What defines X internally?})]
\rev{Decompose an entity into its parts or properties.
While Identify answers ``which one?'' or ``where is it?'' (locating or surfacing a target), Characterize answers ``what kind?'' by describing internal structure (e.g., edge directions within a bundle or the correlation pattern within a cluster).}

  \item[Quantify (\emph{How much of X?})]
  Estimate magnitude; for example, element counts, bundle thicknesses, global density, etc.

  \item[Compare (\emph{How does X differ from Y?})]
  Assess similarities or differences between entities at the same scope level. 
  At the bundle level, this includes comparing thickness or detecting splits and merges; at the multi-view level, comparing bundled views across datasets, time steps, or parameter settings.

  \item[Assess (\emph{Is X faithful?})]
  Evaluate whether the bundled representation truthfully reflects the underlying data. This covers distortion (how far bundling displaced elements), artifact detection (whether perceived patterns are genuine), and parameter sensitivity. This action is specific to bundling.
\end{description}

\noindent\textbf{Compound tasks.}
Some analytical operations span multiple cells.
For example, determining that a particular node bridges two clusters requires first locating the clusters (\sBundle~$\times$~Identify), then tracing individual edges from the node to verify it connects to both (\sElement~$\times$~Identify).
Following Oddo et al.~\cite{Oddo2026}, we treat these as chained sequences of atomic tasks rather than adding combinations.

\subsection{Representation Type as Instantiation}
\label{sec:reptype}

Each Scope~$\times$~Action cell defines an abstract task.
The representation type determines concrete semantics, because the same abstract task means different things for node-link diagrams, trail sets, and PCPs.
Consider three examples.

\sElement~$\times$~Verify asks whether a specific connection exists.
In a node-link diagram, this means checking whether two nodes share an edge.
In a trail set, it means checking whether any trajectory links two regions.
In a PCP, it means checking whether a record with a particular value combination exists.
All three are existence checks, but the data semantics differ fundamentally.

\sBundle~$\times$~Characterize asks what defines a bundle internally.
In a node-link diagram, this means examining the directional mix of edges within a bundle.
In a trail set, it means describing the mix of short and long trips that defines an area.
In a PCP, it means determining whether the variables within a cluster correlate positively or negatively.
This last instantiation is specific to PCPs and has no meaningful counterpart in the other two types.

\sGlobal~$\times$~Assess asks whether bundling is faithful.
In a node-link diagram, this concerns edge displacement from the original layout.
In a trail set, the question is whether bundling shifted routes from their true spatial paths, a stronger requirement because the geometry carries real-world meaning.
In a PCP, distortion occurs only in the inter-axis space, leaving values at the axes intact.

These examples illustrate why a single abstract taxonomy requires representation-specific interpretation.
We flag representation-specific instantiations in our coded corpus rather than splitting them into separate tasks.

\section{Corpus Analysis}
\label{sec:corpus}

\begin{table}[t]
\centering
\caption{Distribution of task mentions across the Scope~$\times$~Action matrix (49~papers, 444~mentions). Gray cells mark tasks hindered by bundling; empty cells are structurally impossible and omitted.}
\label{tab:distribution}
\fontsize{7}{7}\selectfont
\setlength{\tabcolsep}{2pt}
\renewcommand{\arraystretch}{1.0}
\begin{tabular}{l cccccc r}
 & \rotatebox{35}{\textbf{Verify}} & \rotatebox{35}{\textbf{Identify}} & \rotatebox{35}{\textbf{Characterize}} & \rotatebox{35}{\textbf{Quantify}} & \rotatebox{35}{\textbf{Compare}} & \rotatebox{35}{\textbf{Assess}} & \\
\midrule
\textcolor{scopeElement}{\textbf{Element}}
  & 6 & 52 & \cellcolor{black!8}{0} & 1 & \cellcolor{black!8}{0} & 17 & 76 \\
\textcolor{scopeBundle}{\textbf{Bundle}}
  & 15 & \textbf{80} & 36 & 26 & 20 & 9 & \textbf{186} \\
\textcolor{scopeGlobal}{\textbf{Global}}
  & 1 & 51 & 5 & 7 & & \textbf{72} & 136 \\
\textcolor{scopeMulti}{\textbf{Multi-view}}
  & & & & & 43 & 3 & 46 \\
\midrule
 & 22 & 183 & 41 & 34 & 63 & 101 & 444 \\
\bottomrule
\end{tabular}
\end{table}

\rev{Of the 102 papers, 49 contained at least one bundling-attributed task and form the basis of the taxonomy; the remainder either address bundling algorithmically without discussing tasks, or use bundling so peripherally that no task passed the attribution filter.}
By representation type, 322 mentions (73\%) concern node-link diagrams, 78 (16\%) trail sets, and 44 (11\%) PCPs.
The vast majority of mentions are explicit (92\%); only 36 are implicit, a ratio that likely reflects both the prevalence of explicit task claims in the bundling literature and a limitation of LLM-based screening, which may undercount implicit mentions.
\cref{tab:distribution} shows the per-cell distribution and we highlight the key patterns below.

Identify is the dominant action with 183 mentions (41\%).
\rev{At the bundle level, 31 of 49 papers claim their method ``shows major flows,'' ``makes clusters visible,'' or ``surfaces connection groups,'' all instances of \sBundle~$\times$~Identify (locating which aggregates are present, not describing their internal composition).}
At the global level, 32 papers frame bundling as ``providing an overview'' or ``reducing clutter to reveal structure'' (\sGlobal~$\times$~Identify).

Assess is the second-largest action (101 mentions, 23\%), but its distribution is uneven.
\sGlobal~$\times$~Assess accounts for 72 mentions across 36 papers, covering distortion, parameter sensitivity, and the risk of false patterns.
In contrast, \sBundle~$\times$~Assess (9 mentions, 8 papers) is notably sparse: few papers ask whether a perceived bundle is genuine or an artifact, despite this being a known risk~\cite{LhuillierHT17}.

The \sBundle row is the richest scope with 186 mentions spread across all six actions, confirming that bundles function as first-class perceptual objects with their own task vocabulary.
The \sMulti row is dominated by Compare (43 of 46 mentions, 31 papers), predominantly in trail-set contexts.

Two cells received zero mentions: \sElement~$\times$~Characterize and \sElement~$\times$~Compare.
As discussed in \cref{sec:scope}, these are coherent tasks that bundling hinders rather than empty cells.
Their absence from the corpus is consistent with the enabling/disabling duality and is reinforced by interaction techniques explicitly designed to recover element-level access~\cite{WongCG03,HurterET11}.

\section{Discussion}
\label{sec:discussion}

\paragraph{Bundling enables and disables tasks simultaneously.}
The corpus confirms a duality not captured by prior frameworks: bundling \emph{enables} tasks at the bundle and \sGlobal levels by introducing perceptual aggregates, while it \emph{disables} tasks at the element level by merging individual elements.
Prior task discussions~\cite{LhuillierHT17} list capabilities that strong bundling effectively disables (e.g., precise attribute reading) while omitting tasks that bundling clearly enables (e.g., global faithfulness assessment, bundle-thickness comparison).
A taxonomy that does not represent both directions will overstate what bundling supports.
\rev{This dual structure makes the taxonomy usable in three ways: descriptively (classifying the tasks a paper targets), generatively (surfacing gaps), and evaluatively (matching quality metrics to intended tasks).}

\paragraph{Evaluation gaps.}
\rev{The empty and sparse cells translate into concrete research questions for future work.
\emph{RQ1:} Can viewers reliably distinguish a genuine bundle from a bundling artifact? \sBundle~$\times$~Assess appears in only 8 papers despite this known risk.
\emph{RQ2:} How do bundling parameters affect which patterns are visible across views? Systematic \sMulti comparison of parameter settings is almost absent.
\emph{RQ3:} Do tasks transfer to trail and PCP bundling as readily as to node-link diagrams? The 73/16/11 split reflects a literature bias.
\emph{RQ4:} Which quality metrics predict performance on which tasks? Most papers report metrics without specifying the tasks.
}

\paragraph{Limitations.}
Our coding inherits the literature's biases: node-link papers dominate (73\% of mentions), and PCP bundling is underrepresented.
The LLM-assisted screening may also undercount implicit task mentions, as suggested by the 92\%/8\% explicit/implicit ratio.
\rev{We also note the absence of a formal inter-rater agreement coefficient as a limitation of our coding procedure.}
The taxonomy is grounded in three established bundling settings; some variants (e.g., 3D bundling, bundling on non-Euclidean layouts) may require additional instantiations.

\section{Conclusion}
\label{sec:conclusion}

We present a task taxonomy for edge and trail bundling, derived from a systematic analysis of 102~papers and organized as a Scope~$\times$~Action matrix.
By extending the analysis beyond node-link diagrams to include trail sets and parallel coordinates, and by recognizing bundles as first-class perceptual objects, the taxonomy provides a structured vocabulary for reasoning about what can be done with bundled visualizations.
The two-dimensional structure enables descriptive use (classifying existing tasks), generative use (identifying empty cells as evaluation gaps), and evaluative use (connecting bundling quality metrics to the tasks they serve).
We invite the community to use the taxonomy when designing bundling evaluations, selecting quality metrics, and reporting task coverage in future bundling papers.

\section*{Supplemental Materials}
\rev{
The supplemental material provides the full coded corpus underlying our
taxonomy. It contains the complete set of 444 bundling-attributed task
mentions across the 49 papers, each coded by scope, action, and representation
type, and annotated with the source paper's original terminology and the
evidence attributing the task to bundling. Both a machine-readable table (one
row per mention, with paper metadata) and a typeset version of the corpus are
included, along with a cross-tabulation of mentions by scope, action, and
representation type. We further provide the screening materials used during
coding: the coding scheme with synonym lists mapping literature phrasings to
our canonical labels, the bundling attribution filter with worked examples, and
the scripts used to run the first-pass LLM screening over the paper collection.
All supplemental material is available on \osf.
}

\bibliographystyle{abbrv-doi}
\bibliography{bibliography}

@article{HurterET11,
  author    = {Christophe Hurter and Ozan Ersoy and Alexandru Telea},
  title     = {{MoleView}: An Attribute and Structure-Based Semantic Lens
               for Large Element-Based Plots},
  journal   = {IEEE Transactions on Visualization and Computer Graphics},
  volume    = {17},
  number    = {12},
  pages     = {2600--2609},
  year      = {2011},
  doi       = {10.1109/TVCG.2011.223}
}

@article{VehlowBW17,
  author       = {Corinna Vehlow and
                  Fabian Beck and
                  Daniel Weiskopf},
  title        = {Visualizing Group Structures in Graphs: {A} Survey},
  journal      = {Comput. Graph. Forum},
  volume       = {36},
  number       = {6},
  pages        = {201--225},
  year         = {2017},
  url          = {https://doi.org/10.1111/cgf.12872},
  doi          = {10.1111/CGF.12872},
  bibsource    = {dblp computer science bibliography, https://dblp.org}
}

@inproceedings{WongCG03,
  author    = {Nelson Wong and Sheelagh Carpendale and Saul Greenberg},
  title     = {EdgeLens: An Interactive Method for Managing Edge
               Congestion in Graphs},
  booktitle = {IEEE Symposium on Information Visualization (InfoVis)},
  pages     = {51--58},
  year      = {2003},
  doi       = {10.1109/INFVIS.2003.1249008}
}

@article{TaiBXS24,
  author    = {Robert H. Tai and Lillian R. Bentley and Xin Xia
               and Jason M. Sitt and Sarah C. Fankhauser
               and Ana M. Chicas-Mosier and Barnas G. Monteith},
  title     = {An Examination of the Use of Large Language Models
               to Aid Analysis of Textual Data},
  journal   = {International Journal of Qualitative Methods},
  volume    = {23},
  year      = {2024},
  doi       = {10.1177/16094069241231168}
}

@article{HeinrichW15,
  author       = {Julian Heinrich and
                  Daniel Weiskopf},
  title        = {Parallel Coordinates for Multidimensional Data Visualization: Basic
                  Concepts},
  journal      = {Comput. Sci. Eng.},
  volume       = {17},
  number       = {3},
  pages        = {70--76},
  year         = {2015},
  url          = {https://doi.org/10.1109/MCSE.2015.55},
  doi          = {10.1109/MCSE.2015.55},
  bibsource    = {dblp computer science bibliography, https://dblp.org}
}

@article{HurterTC09,
  author       = {Christophe Hurter and
                  Benjamin Tissoires and
                  St{\'{e}}phane Conversy},
  title        = {FromDaDy: Spreading Aircraft Trajectories Across Views to Support
                  Iterative Queries},
  journal      = {{IEEE} Trans. Vis. Comput. Graph.},
  volume       = {15},
  number       = {6},
  pages        = {1017--1024},
  year         = {2009},
  url          = {https://doi.org/10.1109/TVCG.2009.145},
  doi          = {10.1109/TVCG.2009.145},
  bibsource    = {dblp computer science bibliography, https://dblp.org}
}

@article{epb,
  author       = {Markus Wallinger and
                  Daniel Archambault and
                  David Auber and
                  Martin N{\"{o}}llenburg and
                  Jaakko Peltonen},
  title        = {Edge-Path Bundling: {A} Less Ambiguous Edge Bundling Approach},
  journal      = {{IEEE} Trans. Vis. Comput. Graph.},
  volume       = {28},
  number       = {1},
  pages        = {313--323},
  year         = {2022},
  urlOld          = {https://doi.org/10.1109/TVCG.2021.3114795},
  doi          = {10.1109/TVCG.2021.3114795},
  bibsource    = {dblp computer science bibliography, https://dblp.org}
}

@INPROCEEDINGS{lhuiller_2017,
  author       = {Antoine Lhuillier and
                  Christophe Hurter and
                  Alexandru C. Telea},
  editor_old       = {Daniel Weiskopf and
                  Yingcai Wu and
                  Tim Dwyer},
  title        = {{FFTEB:} Edge bundling of huge graphs by the Fast Fourier Transform},
  booktitle    = {2017 {IEEE} Pacific Visualization Symposium, PacificVis 2017, Seoul,
                  South Korea, April 18-21, 2017},
  pages        = {190--199},
  publisher    = {{IEEE} Computer Society},
  year         = {2017},
  urlOld          = {https://doi.org/10.1109/PACIFICVIS.2017.8031594},
  doi          = {10.1109/PACIFICVIS.2017.8031594},
  bibsource    = {dblp computer science bibliography, https://dblp.org}
}

@article{lambert_2010,
  author       = {Antoine Lambert and
                  Romain Bourqui and
                  David Auber},
  title        = {Winding Roads: Routing edges into bundles},
  journal      = {Comput. Graph. Forum},
  volume       = {29},
  number       = {3},
  pages        = {853--862},
  year         = {2010},
  urlOld          = {https://doi.org/10.1111/j.1467-8659.2009.01700.x},
  doi          = {10.1111/J.1467-8659.2009.01700.X},
  bibsource    = {dblp computer science bibliography, https://dblp.org}
}

@article{spanepb,
  author       = {Markus Wallinger and
                  Daniel Archambault and
                  David Auber and
                  Martin N{\"{o}}llenburg and
                  Jaakko Peltonen},
  title        = {Faster Edge-Path Bundling through Graph Spanners},
  journal      = {Comput. Graph. Forum},
  volume       = {42},
  number       = {6},
  year         = {2023},
  pages        = {e14789},
  urlOld          = {https://doi.org/10.1111/cgf.14789},
  doi          = {10.1111/CGF.14789},
  bibsource    = {dblp computer science bibliography, https://dblp.org}
}

@inproceedings{Saga16,
  author       = {Ryosuke Saga},
  editor_old       = {Tobias Isenberg and
                  Filip Sadlo},
  title        = {Quantitative Evaluation for Edge Bundling Based on Structural Aesthetics},
  booktitle    = {18th Eurographics Conference on Visualization, EuroVis 2016 - Posters,
                  Groningen, The Netherlands, June 6-10, 2016},
  pages        = {17--19},
  publisher    = {Eurographics Association},
  year         = {2016},
  urlOld          = {https://doi.org/10.2312/eurp.20161131},
  doi          = {10.2312/EURP.20161131},
  bibsource    = {dblp computer science bibliography, https://dblp.org}
}

@inproceedings{ArchambaultLNPT24,
  author       = {Daniel Archambault and
                  Giuseppe Liotta and
                  Martin N{\"{o}}llenburg and
                  Tommaso Piselli and
                  Alessandra Tappini and
                  Markus Wallinger},
  @editor_old       = {Stefan Felsner and
                  Karsten Klein},
  title        = {Bundling-Aware Graph Drawing},
  booktitle    = {32nd International Symposium on Graph Drawing and Network Visualization,
                  {GD} 2024, September 18-20, 2024, Vienna, Austria},
  series       = {LIPIcs},
  volume       = {320},
  pages        = {15:1--15:19},
  publisher    = {Schloss Dagstuhl - Leibniz-Zentrum f{\"{u}}r Informatik},
  year         = {2024},
  urlOld          = {https://doi.org/10.4230/LIPIcs.GD.2024.15},
  doi          = {10.4230/LIPICS.GD.2024.15},
  timestamp    = {Mon, 28 Oct 2024 16:46:06 +0100},
  biburl       = {https://dblp.org/rec/conf/gd/ArchambaultLNPT24.bib},
  bibsource    = {dblp computer science bibliography, https://dblp.org}
}

@article{WallingerPTALN25,
  author       = {Markus Wallinger and
                  Tommaso Piselli and
                  Alessandra Tappini and
                  Daniel Archambault and
                  Giuseppe Liotta and
                  Martin N{\"{o}}llenburg},
  title        = {Bundling-Aware Graph Drawing Revisited},
  journal      = {{IEEE} Trans. Vis. Comput. Graph.},
  volume       = {31},
  number       = {12},
  pages        = {10828--10839},
  year         = {2025},
  urlOld          = {https://doi.org/10.1109/TVCG.2025.3616583},
  doi          = {10.1109/TVCG.2025.3616583},
  bibsource    = {dblp computer science bibliography, https://dblp.org}
}

@inproceedings{WallingerARPA25,
  author       = {Markus Wallinger and
                  Osman Akbulut and
                  Kabir Ahmed Rufai and
                  Helen C. Purchase and
                  Daniel Archambault},
  editor_old       = {Naomi Yamashita and
                  Vanessa Evers and
                  Koji Yatani and
                  Sharon Xianghua Ding and
                  Bongshin Lee and
                  Marshini Chetty and
                  Phoebe O. Toups Dugas},
  title        = {How Do People Perceive Bundling? An Experiment},
  booktitle    = {Proceedings of the 2025 {CHI} Conference on Human Factors in Computing
                  Systems, {CHI} 2025, Yokohama, Japan, 26 April 2025- 1 May 2025},
  pages        = {1113:1--1113:14},
  publisher    = {{ACM}},
  year         = {2025},
  doi          = {10.1145/3706598.3713444},
  bibsource    = {dblp computer science bibliography, https://dblp.org}
}

@inproceedings{EllisD06,
  author       = {Geoffrey P. Ellis and
                  Alan J. Dix},
  title        = {The plot, the clutter, the sampling and its lens: occlusion measures
                  for automatic clutter reduction},
  booktitle    = {Proceedings of the working conference on Advanced visual interfaces,
                  {AVI} 2006, Venezia, Italy, May 23-26, 2006},
  pages        = {266--269},
  year         = {2006},
  crossref     = {DBLP:conf/avi/2006},
  url_old          = {https://doi.org/10.1145/1133265.1133318},
  doi          = {10.1145/1133265.1133318},
  bibsource    = {dblp computer science bibliography, https://dblp.org}
}

@proceedings{DBLP:conf/avi/2006,
  editor_old       = {Augusto Celentano},
  title        = {Proceedings of the working conference on Advanced visual interfaces,
                  {AVI} 2006, Venezia, Italy, May 23-26, 2006},
  publisher    = {{ACM} Press},
  year         = {2006},
  url_old          = {https://doi.org/10.1145/1133265},
  doi          = {10.1145/1133265},
  isbn         = {1-59593-353-0},
  bibsource    = {dblp computer science bibliography, https://dblp.org}
}

@article{EllisD07,
  author       = {Geoffrey P. Ellis and
                  Alan J. Dix},
  title        = {A Taxonomy of Clutter Reduction for Information Visualisation},
  journal      = {{IEEE} Trans. Vis. Comput. Graph.},
  volume       = {13},
  number       = {6},
  pages        = {1216--1223},
  year         = {2007},
  url          = {https://doi.org/10.1109/TVCG.2007.70535},
  doi          = {10.1109/TVCG.2007.70535},
  bibsource    = {dblp computer science bibliography, https://dblp.org}
}

@article{BrehmerM13,
  author       = {Matthew Brehmer and
                  Tamara Munzner},
  title        = {A Multi-Level Typology of Abstract Visualization Tasks},
  journal      = {{IEEE} Trans. Vis. Comput. Graph.},
  volume       = {19},
  number       = {12},
  pages        = {2376--2385},
  year         = {2013},
  url          = {https://doi.org/10.1109/TVCG.2013.124},
  doi          = {10.1109/TVCG.2013.124},
  bibsource    = {dblp computer science bibliography, https://dblp.org}
}

@inproceedings{LeePPFH06,
  author       = {Bongshin Lee and
                  Catherine Plaisant and
                  Cynthia Sims Parr and
                  Jean{-}Daniel Fekete and
                  Nathalie Henry},
  title        = {Task taxonomy for graph visualization},
  booktitle    = {Proceedings of the 2006 {AVI} Workshop on BEyond time and errors:
                  novel evaluation methods for information visualization, {BELIV} 2006,
                  Venice, Italy, May 23, 2006},
  pages        = {1--5},
  year         = {2006},
  crossref     = {DBLP:conf/avi/2006beliv},
  url          = {https://doi.org/10.1145/1168149.1168168},
  doi          = {10.1145/1168149.1168168},
  bibsource    = {dblp computer science bibliography, https://dblp.org}
}

@proceedings{DBLP:conf/avi/2006beliv,
  editor       = {Enrico Bertini and
                  Catherine Plaisant and
                  Giuseppe Santucci},
  title        = {Proceedings of the 2006 {AVI} Workshop on BEyond time and errors:
                  novel evaluation methods for information visualization, {BELIV} 2006,
                  Venice, Italy, May 23, 2006},
  publisher    = {{ACM} Press},
  year         = {2006},
  isbn         = {1-59593-562-2},
  bibsource    = {dblp computer science bibliography, https://dblp.org}
}

@article{BachRHMD17,
  author       = {Benjamin Bach and
                  Nathalie Henry Riche and
                  Christophe Hurter and
                  Kim Marriott and
                  Tim Dwyer},
  title        = {Towards Unambiguous Edge Bundling: Investigating Confluent Drawings
                  for Network Visualization},
  journal      = {{IEEE} Trans. Vis. Comput. Graph.},
  volume       = {23},
  number       = {1},
  pages        = {541--550},
  year         = {2017},
  url          = {https://doi.org/10.1109/TVCG.2016.2598958},
  doi          = {10.1109/TVCG.2016.2598958},
  bibsource    = {dblp computer science bibliography, https://dblp.org}
}

@article{CuiZQWL08,
  author       = {Weiwei Cui and
                  Hong Zhou and
                  Huamin Qu and
                  Pak Chung Wong and
                  Xiaoming Li},
  title        = {Geometry-Based Edge Clustering for Graph Visualization},
  journal      = {{IEEE} Trans. Vis. Comput. Graph.},
  volume       = {14},
  number       = {6},
  pages        = {1277--1284},
  year         = {2008},
  url          = {https://doi.org/10.1109/TVCG.2008.135},
  doi          = {10.1109/TVCG.2008.135},
  bibsource    = {dblp computer science bibliography, https://dblp.org}
}

@article{LhuillierHT17,
  author       = {Antoine Lhuillier and
                  Christophe Hurter and
                  Alexandru C. Telea},
  title        = {State of the Art in Edge and Trail Bundling Techniques},
  journal      = {Comput. Graph. Forum},
  volume       = {36},
  number       = {3},
  pages        = {619--645},
  year         = {2017},
  url          = {https://doi.org/10.1111/cgf.13213},
  doi          = {10.1111/CGF.13213},
  bibsource    = {dblp computer science bibliography, https://dblp.org}
}

@article{ZwanCT16,
  author       = {Matthew van der Zwan and
                  Valeriu Codreanu and
                  Alexandru C. Telea},
  title        = {CUBu: Universal Real-Time Bundling for Large Graphs},
  journal      = {{IEEE} Trans. Vis. Comput. Graph.},
  volume       = {22},
  number       = {12},
  pages        = {2550--2563},
  year         = {2016},
  url          = {https://doi.org/10.1109/TVCG.2016.2515611},
  doi          = {10.1109/TVCG.2016.2515611},
  bibsource    = {dblp computer science bibliography, https://dblp.org}
}

@article{HurterET12,
  author       = {Christophe Hurter and
                  Ozan Ersoy and
                  Alexandru C. Telea},
  title        = {Graph Bundling by Kernel Density Estimation},
  journal      = {Comput. Graph. Forum},
  volume       = {31},
  number       = {3pt1},
  pages        = {865--874},
  year         = {2012},
  url          = {https://doi.org/10.1111/j.1467-8659.2012.03079.x},
  doi          = {10.1111/J.1467-8659.2012.03079.X},
  bibsource    = {dblp computer science bibliography, https://dblp.org}
}

@article{HoltenW09,
  author       = {Danny Holten and
                  Jarke J. van Wijk},
  title        = {Force-Directed Edge Bundling for Graph Visualization},
  journal      = {Comput. Graph. Forum},
  volume       = {28},
  number       = {3},
  pages        = {983--990},
  year         = {2009},
  url          = {https://doi.org/10.1111/j.1467-8659.2009.01450.x},
  doi          = {10.1111/J.1467-8659.2009.01450.X},
  bibsource    = {dblp computer science bibliography, https://dblp.org}
}

@article{Holten06,
  author       = {Danny Holten},
  title        = {Hierarchical Edge Bundles: Visualization of Adjacency Relations in
                  Hierarchical Data},
  journal      = {{IEEE} Trans. Vis. Comput. Graph.},
  volume       = {12},
  number       = {5},
  pages        = {741--748},
  year         = {2006},
  url          = {https://doi.org/10.1109/TVCG.2006.147},
  doi          = {10.1109/TVCG.2006.147},
  bibsource    = {dblp computer science bibliography, https://dblp.org}
}

@article{SaketSKB14,
  author       = {Bahador Saket and
                  Paolo Simonetto and
                  Stephen G. Kobourov and
                  Katy B{\"{o}}rner},
  title        = {Node, Node-Link, and Node-Link-Group Diagrams: An Evaluation},
  journal      = {{IEEE} Trans. Vis. Comput. Graph.},
  volume       = {20},
  number       = {12},
  pages        = {2231--2240},
  year         = {2014},
  url          = {https://doi.org/10.1109/TVCG.2014.2346422},
  doi          = {10.1109/TVCG.2014.2346422},
  bibsource    = {dblp computer science bibliography, https://dblp.org}
}

@inproceedings{PalmasW16,
  author       = {Gregorio Palmas and
                  Tino Weinkauf},
  editor       = {Enrico Bertini and
                  Niklas Elmqvist and
                  Thomas Wischgoll},
  title        = {Space Bundling for Continuous Parallel Coordinates},
  booktitle    = {18th Eurographics Conference on Visualization, EuroVis 2016 - Short
                  Papers, Groningen, The Netherlands, June 6-10, 2016},
  pages        = {61--65},
  publisher    = {Eurographics Association},
  year         = {2016},
  url          = {https://doi.org/10.2312/eurovisshort.20161162},
  doi          = {10.2312/EUROVISSHORT.20161162},
  bibsource    = {dblp computer science bibliography, https://dblp.org}
}

@article{McDonnellM08,
  author       = {Kevin T. McDonnell and
                  Klaus Mueller},
  title        = {Illustrative Parallel Coordinates},
  journal      = {Comput. Graph. Forum},
  volume       = {27},
  number       = {3},
  pages        = {1031--1038},
  year         = {2008},
  url          = {https://doi.org/10.1111/j.1467-8659.2008.01239.x},
  doi          = {10.1111/J.1467-8659.2008.01239.X},
  bibsource    = {dblp computer science bibliography, https://dblp.org}
}

@article{oddo2026,
  title={Visualization Tasks for Unlabeled Graphs},
  author={Oddo, Matt IB and Smith, Ryan and Kobourov, Stephen and Munzner, Tamara},
  journal={{IEEE} Trans. Vis. Comput. Graph.},
  year={2026},
  publisher={IEEE}
}

@article{HenryFM07,
  author       = {Nathalie Henry and
                  Jean{-}Daniel Fekete and
                  Michael J. McGuffin},
  title        = {NodeTrix: a Hybrid Visualization of Social Networks},
  journal      = {{IEEE} Trans. Vis. Comput. Graph.},
  volume       = {13},
  number       = {6},
  pages        = {1302--1309},
  year         = {2007},
  url          = {https://doi.org/10.1109/TVCG.2007.70582},
  doi          = {10.1109/TVCG.2007.70582},
  bibsource    = {dblp computer science bibliography, https://dblp.org}
}

@article{Wallinger2026Metrics,
author = {Wallinger, M. and Miller, J. and Maftei, A. and Kobourov, S. G.},
title = {Quantitative Metrics for Edge Bundling of Network Visualizations},
journal = {Computer Graphics Forum},
volume = {n/a},
number = {n/a},
pages = {e70443},
doi = {https://doi.org/10.1111/cgf.70443},
url = {https://onlinelibrary.wiley.com/doi/abs/10.1111/cgf.70443},
eprint = {https://onlinelibrary.wiley.com/doi/pdf/10.1111/cgf.70443}
}

\end{document}